# A MUON COLLIDER AS A HIGGS FACTORY*

D. Neuffer[#], M. Palmer, Y. Alexahin, Fermilab, Batavia IL 60510, USA, C. Ankenbrandt, Muons, Inc., Batavia IL 60510, USA, J. P. Delahaye, SLAC, Menlo Park, CA 94025 USA

*Abstract*
Because muons connect directly to a standard-model Higgs particle in s-channel production, a muon collider would be an ideal device for precision measurement of the mass and width of a Higgs-like particle, and for further exploration of its production and decay properties. Parameters of a high-precision muon collider are presented and the necessary components and performance are described. An important advantage of the muon collider approach is that the spin precession of the muons will enable energy measurements at extremely high accuracy (dE/E to $10^{-6}$ or better). The collider could be a first step toward a high-luminosity multi-TeV lepton collider, and extensions toward a higher-energy higher-luminosity device are also discussed.

## INTRODUCTION

The CERN ATLAS and CMS collaborations have presented evidence for a Higgs particle at ~126 GeV. The Higgs candidate is consistent with a minimal standard-model Higgs, which has a small production cross-section with a very narrow width. As discussed in Barger et al.,[1] a minimal Higgs could be produced in the s-channel in a muon collider. ($\mu^+ + \mu^- \rightarrow H_0$) The possibility of producing and studying the standard model Higgs at ~100 GeV energy was explored by the Muon Collaboration in 1996-2003[2, 3] and most of that discussion remains valid. In the present paper, we follow more recent studies in muon production, cooling, and acceleration within the MAP Collaboration (Muon Accelerator Program) to obtain scenarios for a Higgs-energy $\mu^+$-$\mu^-$ Collider.[4] Muon spin precession can accurately calibrate the mass and width, and the nearby high cross-section $Z_0$ resonance can be exploited for development and debugging of the facility. Extension of an initial facility toward higher energy and luminosity is discussed.

## OVERVIEW OF A 126 GEV HIGGS $\mu^+$-$\mu^-$ COLLIDER

At 126 GeV, the standard-model Higgs is a narrow resonance with a width of ~4 MeV, and the cross-section for production in $\mu^+$-$\mu^- \rightarrow H_0$ is ~40pb. This is relatively small, but is $(m_\mu/m_e)^2$ larger than for an $e^+$-$e^-$ collider, and a luminosity of $L = 10^{31}$ cm$^{-2}$/s would provide ~4000 $H_0$ / $10^7$ s operational "year". A scan over the Higgs mass with a small-δE collider would resolve that mass and width to high accuracy, much higher than any alternative $H_0$ studies. The initial difficulty will be in isolating the $H_0$ and a scan over a larger initial energy spread will be needed.

An artistic impression of a muon collider is presented in Fig. 1. It consists of a source of high-intensity short proton pulses, a production target with collection of secondary π's, a decay transport, a bunching and cooling channel to capture and cool μ's from π decay into intense bunches, and an accelerator that takes the μ+ and μ- bunches to a collider ring for full-energy collision in an interaction region inside a Detector. Table 1 contains parameters of $\mu^+$-$\mu^-$ Colliders from 126 GeV to 3TeV.

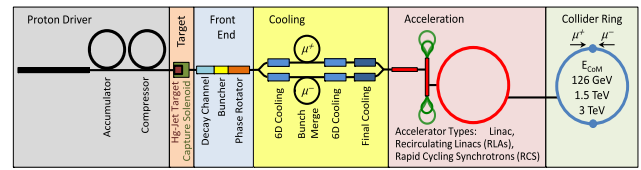

Figure 1: An overview of a μ+-μ- Collider Facility, with collider energies from 126 GeV up to 1.5 and 3 GeV.

*Proton Source, Target and μ Capture Scenario*

The proton source in fig. 1 is based on the Project X 8 GeV linac, upgraded to provide 4MW in 15 Hz pulsed mode.[5] (An alternative configuration using the 3 MW – 3GeV cw linac of Project X phase 2 with collection rings is also being developed.) H- beam from a 15 Hz linac pulse is accumulated over many turns in a storage ring (using charge-exchange injection to H+) and bunched into 4 short bunches, which are extracted two or four at a time to the Front End production target, forming the 30 and 15 Hz cycles used in Table 1.

These bunches are targeted onto a production target producing large number of π's that will decay into μ's. Following the neutrino-factory front-end design [6, 7] this could be a Hg-jet target immersed in a high field solenoid for maximum π-capture, tapering to a lower field transport for π→ μ decay. ~300—200 MHz rf cavities form the μ's into trains of $\mu^+$ and $\mu^-$ bunches, which are phase-energy rotated into equal energy bunches, at which an ionization cooling transport (solenoids +rf + absorbers) initiates the cooling needed for the collider. This Front End is ~150m long.

*Cooling Scenario and Constraints*

The small δE requirement of the Collider implies that the beam must be cooled to minimal longitudinal emittances. The baseline cooling scenario for a Collider starts with bunch trains from the front end and cools them both transversely and longitudinally in a sequence of spiral or helical channels, merges the bunches and further cools the beam both transversely and longitudinally, and then enters a final cooling section with high-field magnets toward minimal transverse emittances, while the longitudinal emittance increases. (see Fig. 2) For the 126 GeV Collider the cooling scenario would be truncated at minimal longitudinal emittance, where $\varepsilon_L$ = ~0.0015m and the transverse emittance $\varepsilon_t$ is ~0.0003m. $\varepsilon_t$ could be further reduced to ~0.00015m.

___________________________________________
* Work supported by DOE under contract DE-AC02-07CH11359.
#neuffer@fnal.gov

At these longitudinal emittances the beam would have an energy width of ~3 MeV, which would be small enough for precision exploration of the Higgs. The effective $\beta^*$ is constrained by the hourglass effect to ~ 3 cm. A larger energy width would enable smaller $\beta^*$ and therefore larger luminosity. (A larger L mode with larger $\delta E$ could be useful in the initial scan for the $H_0$.)

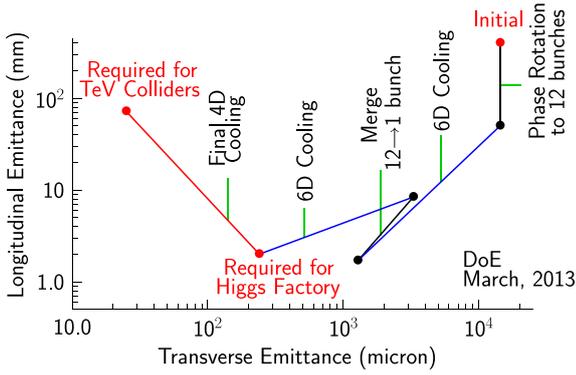

Figure 2. Overview of cooling scenario: Figure A shows a schematic of the cooling beam transports, and B shows the progress of transverse and longitudinal emittances through the system.

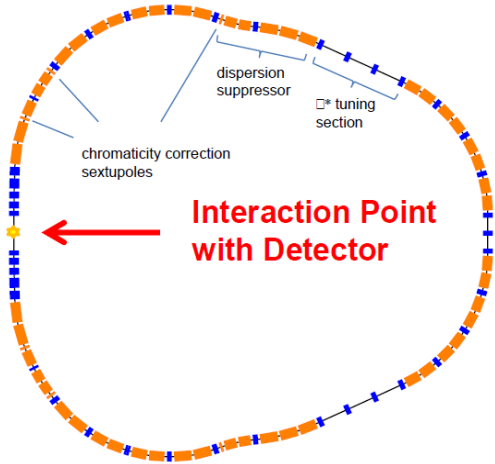

Figure 3. An overview of a 126 GeV $\mu^+$-$\mu^-$ Collider Ring. The circumference is ~300m.

### Acceleration and Collider Ring

Following neutrino factory designs, muon bunches can be accelerated in a linac and a sequence of recirculating linacs (RLA) to 63GeV, where the $\mu^+$ and $\mu^-$ bunches would be inserted into a fixed-field collider ring. A scenario with a 1.8 GeV linac and 2 4.5 pass RLA's (to 7.2 and 62.5 GeV) is a possible extrapolation of the neutrino factory designs, and is used in our initial scenario. [8]

A lattice specific to a 63 GeV/beam collider ring has been developed by Alexahin et al.[9] The 300m ring is dominated by matching into the interaction region, integrated with bending, correction and tuning functions. (see fig. 3) The lattice has been designed to operate in a small $\delta E$ mode to maximize production on resonance and a larger $\delta E$ mode with higher luminosity adapted to the energy search. Beam-beam tune shifts are modest.

## ENERGY DETERMINATION BY SPIN TRACKING

Raja and Tollestrup noted that the energy of the beams can be measured to high accuracy by tracking the precession of the decay electron energies.[10] While stored, the muons continuously decay following $\mu \rightarrow e + \nu_\mu + \bar{\nu}_e$, at ~$10^7$ decays per m, and the electrons and positrons from the decay have a mean energy dependent on the polarization of the muons. That polarization $\hat{P}$ will precess as the beam rotates around the ring and that precession will modulate the mean energy of decay electrons, and therefore the signal at a detector capturing those decays. In the present scenario the $\mu$ beams are created with a small polarization (~10—20% from a bias toward capture of forward $\pi \rightarrow \mu$ decays) and that polarization should be substantially maintained through the cooling and acceleration systems. The mean energy from decay electrons is:

$$\langle E(t) \rangle = \left\langle Ne^{-\alpha t}\left(\tfrac{7}{20} E_\mu \left(1 + \tfrac{\beta}{7}\hat{P}\cos[\omega t + \phi]\right)\right)\right\rangle,$$

where $N$ is the initial number of $\mu$'s, $E_\mu$ is the $\mu$ energy, $\alpha$ is the decay parameter, $\beta = v/c$, $P$ is the polarization, $\phi$ is a phase, $t$ is time in turn numbers and $\omega = 2\pi\lambda\left(\tfrac{g-2}{2}\right) \cong 2\pi \cdot 0.7$ is the precession frequency that depends on the muon beam energy. A detector capturing a significant number of decay electrons will have a signal modulated by that precession frequency which can be measured to very high accuracy, obtaining an energy measurement to the ~$10^{-6}$ level (corresponding to 0.1 MeV), or better.

### Collider at the $Z_0$: "Training Wheels" for the Higgs Factory

Initial operation of a collider at a small-$\delta E$ $H_0$ appears quite daunting, particularly since initial luminosities will be less than desired. However the 126 GeV Higgs is quite close to the 91.2 GeV $Z_0$, where the production cross section is almost 1000 times larger (~30nb), and a luminosity of only ~$10^{27}$ would see $Z_0$ production events. (see Fig. 4) We propose to initially operate and debug the facility at 91.2 GeV. The large cross section nearly guarantees the existence of non-background events in early operation and the difficult task of separating signal from backgrounds can be initiated at relatively easy parameters. The energy measurement technique would be debugged at a well-known value, and a sweep of a small δE Collider over the 2.5 GeV width of the $Z_0$ would provide valuable information on the collider operation and nontrivial information on $Z_0$ properties. The acceleration and storage will then be increased from 45.6 to 63 GeV, and the scan for the Higgs will begin.

Although the $Z_0$ is well known, the comparison of $\mu^+$-$\mu^- \rightarrow Z_0$ with $e^+$-$e^-$ is of some interest, and the spin precession measurement of the energy and width could be more accurate than existing values ( now at δE ~ 2 MeV).

Table 1: Parameters of $\mu^+$-$\mu^-$ Colliders from 126 GeV to 3 TeV.

| Parameter | Unit | Higgs I | Higgs II | 1.5TeV | High Energy |
|---|---|---|---|---|---|
| Collision Energy | GeV | 126 | 126 | 1500 | 3000 |
| Beam energy | GeV | 63 | 63 | 750 | 1500 |
| Average luminosity | $10^{31}$/cm$^2$/s | 1.7 | 8.0 | 1250 | 4400 |
| Collision energy spread | MeV | 3 | 4 | 750 | 1500 |
| Circumference, C | m | 300 | 300 | 2500 | 4450 |
| Number of IPs | - | 1 | 1 | 2 | 2 |
| β* | cm | 3.3 | 1.7 | 1.0 | 0.5 |
| Number of muons / bunch | $10^{12}$ | 2 | 4 | 2 | 2 |
| Number of bunches / beam | - | 1 | 1 | 1 | 1 |
| Beam energy spread | % | 0.003 | 0.004 | 0.1 | 0.1 |
| Normalized emittance, $\varepsilon_{\perp N}$ | mm·rad | 0.4 | 0.2 | 0.025 | 0.025 |
| Longitudinal emittance, $\varepsilon_{\parallel N}$ | mm | 1.0 | 1.5 | 70 | 70 |
| Bunch length, $\sigma_s$ | cm | 5.6 | 6.3 | 1.0 | 0.5 |
| Beam size at IP, r.m.s. | mm | 0.15 | 0.075 | 0.006 | 0.003 |
| Beam size in IR quads, r.m.s. | cm | 4 | 4 | 1.4 | 1.4 |
| Beam-beam parameter | - | 0.005 | 0.02 | 0.09 | 0.09 |
| Repetition rate | Hz | 30 | 15 | 15 | 12 |
| Proton driver power | MW | 4 | 4 | 4 | 4 |

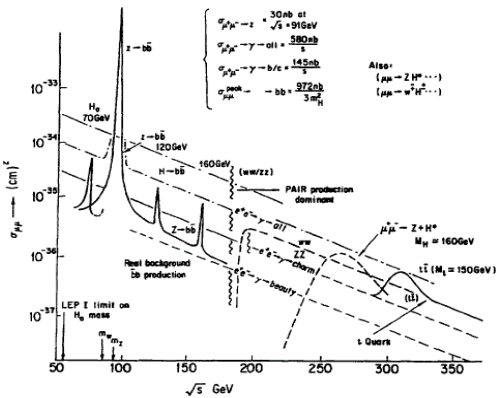

Figure 4. An overview of σ(μμ→??) at E = 50 to 350 GeV, showing the Z peak, possible H0 results, and other known effects. (from ref. 11)

## LUMINOSITY AND ENERGY UPGRADES

The Collider can be optimized and improved toward higher luminosity. More cooling could reduce transverse emittance. Stronger focusing could reduce β* by a factor of 2—4. For higher luminosity, the 4 bunches from the accumulator can be combined to hit the target at the same time, reducing the cycling frequency to 15 Hz, but increasing the μ's /cycle by a factor of 2. These and other upgrades could increase L to ~$10^{32}$, (see table 1) but not much larger in a small-δE mode, unless the proton source power is increased, or beam cooling is greatly improved.

New physics beyond 126 GeV can be explored by higher-energy $\mu^+$-$\mu^-$ Colliders, and the Higgs collider can be extended to a high-luminosity, high-energy facility by the addition of final cooling, more acceleration, and a larger collider ring. Acceleration to higher energies may use a very-rapid-cycling synchrotron scenario. At higher energies the resonance widths are larger, and the beams could have larger $\varepsilon_L$, and would therefore be cooled transversely by another order of magnitude. (see fig. 3). That and adiabatic damping would increase luminosity to $> 10^{34}$ for a multi-TeV Collider. Table 1 displays potential energy and luminosity upgrades up to 3 TeV. Higher energy extensions are being explored by the MAP collaboration.

## REFERENCES

[1] V. Barger, M. S. Berger, J. F. Gunion and T. Han, *Physics Reports* **286**, 1-51 (1997).
[2] C. Ankenbrandt et al., *Physical Review STAB* **2**, 081001 (1999).
[3] M. M. Alsharo'a et al., *Physical Review STAB* **6**, 081001 (2003).
[4] D. Neuffer, "The First Muon Collider – 125 GeV Higgs Factory?" Proc. AAC2012, A.I. P. Conf. Proc. 1507, 849 (2012).
[5] Y. Alexahin and D. Neuffer, "Design of Accumulator and Compressor Rings for the Project-X Based Proton Driver", Proc. IPAC2012, New Orleans, LA TUPPC043, p. 1260 (2012).
[6] D. Neuffer et al., "IDR Neutrino Factory Front End and Variations", paper TUPPD006, Proc. IPAC2012 p. 1416 (2012).
[7] M. Appollonio et al., "Accelerator Concept for Future Neutrino Facilities", RAL-TR-2007-23, *JINST* **4** P07001 (2009).
[8] R. Palmer et al., "Scheme for Ionization Cooling for a Muon Collider", Proc. NuFACT08, Valencia, Spain, July 2008.
[9] A. Zlobin et al. paper TUPFI061, Proc. IPAC13, Shanghai, China (2013).
[10] R. Raja and A. Tollestrup, *Physical Review D* **58**, 013005 (1998).
[11] D. Cline, AIP Conf. Proc. 352, A. I. P., Woodbury, NY, 1996, pp. 3-6.